# Carrier statistics and quantum capacitance effects on mobility extraction in two-dimensional crystal semiconductor field-effect transistors

*Nan Ma[*] and Debdeep Jena*

Department of Electrical Engineering, University of Notre Dame,

Notre Dame, Indiana 46556, United States

**Abstract**: In this work, the consequence of the high band-edge density of states on the carrier statistics and quantum capacitance in transition metal dichalcogenide two-dimensional semiconductor devices is explored. The study questions the validity of commonly used expressions for extracting carrier densities and field-effect mobilities from the transfer characteristics of transistors with such channel materials. By comparison to experimental data, a new method for the accurate extraction of carrier densities and mobilities is outlined. The work thus highlights a fundamental difference between these materials and traditional semiconductors that must be considered in future experimental measurements.

---

[*] nma@nd.edu

Two-dimensional (2D) semiconductor crystals, such as the transition metal dichalcogenides (TMDs), are attractive for atomically thin field-effect transistors (FETs) with no broken bonds.[1,2] Coupling the electrostatic advantages with appreciable transport properties in these materials indicates a possibility of high-performance device applications.[3-5] As with graphene, the weak interlayer coupling allows TMD individual layers to be isolated and studied. In contrast to graphene, however, the large energy bandgap of 2D semiconductors enables high on/off current ratio FETs.[6,7] Most properties of interest in FETs originate in the statistics of electrons in the conduction band (CB) and holes in the valence band (VB). The electrostatic field-effect control of these mobile carriers by gates, and their transport properties completely determine the device characteristics. Consequently, the methods employed to extract various parameters from the device characteristics, such as the carrier density and mobility must pay careful attention to the carrier statistics and its link with transport.[8] This has not been done for 2D crystal semiconductors yet. This work presents these fundamental results and identifies a number of errors that arise if the carrier statistics effects are neglected, and provides methods for accurate parameter extractions.

For a single-gate FET with a single-layer (SL) 2D semiconductor channel, the electron density in the channel is usually written as:[9]

$$n_{ox} = C_{ox}(V_{gs} - V_{th})/q, \qquad (1)$$

where $C_{ox} = \varepsilon_{ox}/t_{ox}$ is the gate oxide capacitance per unit area, and $\varepsilon_{ox}$ and $t_{ox}$ are the dielectric constant and thickness of the dielectric layer respectively. $V_{gs}$ is the gate voltage, $V_{th}$ the threshold voltage, and $q$ is the electron charge. The gate capacitance $C_{tot}$ in an FET is the total capacitance of $C_q$ and $C_{ox}$ connected in series, where $C_q$ is the quantum capacitance of the channel.[8,10,11] $C_{tot}$ is dominated by the smaller capacitance. Thus Eq. (1) is only valid when $C_q \gg C_{ox}$. However, for devices with thin high-κ gate dielectrics, or for nondegenerate carrier statistics when the Fermi level is located deep inside the bandgap, $C_q$ can be comparable, or even lower than $C_{ox}$, making Eq. (1) no longer valid. This calls for re-analyzing the carrier statistics and quantum capacitance for TMD channels.

The *E-k* dispersion of mobile carrier states in 2D semiconductors near the bottom of the CB and the top of the VB in the first Brillouin zone is accurately captured by the parabolic approximation: $E(k) = \hbar^2 k^2 / 2m^*$, where $\hbar$ is the reduced Planck constant, $m^*$ is the band-edge effective mass, and $k = \sqrt{k_x^2 + k_y^2}$ is the in-plane 2D wave vector. The band-edge density of states (DOS) is then given by $g(E) = g_s g_v m^* / 2\pi\hbar^2$, where $g_s$ and $g_v$ are the spin and valley degeneracy factors respectively. The 2D carrier densities in the CB and VB are accurately decribed as $n = \int_{E_c}^{\infty} g(E) f(E) dE$ and $p = \int_{-\infty}^{E_v} g(E) [1 - f(E)] dE$, where $E_c$ and $E_v$ are the band-edge energies of the CB and VB respectively. The occupation probability is the Fermi-Dirac distribution $f(E) = 1 / \{1 + \exp[(E - E_f)/k_B T]\}$, with $k_B$ the Boltzmann constant, $T$ the absolute temperature, and $E_f$ the Fermi level. From above equations, the electron density in the CB is $n = g_{2D} k_B T \ln\{1 + \exp[(E_f - E_c)/k_B T]\}$ and the hole density in the VB is $p = g_{2D} k_B T \ln\{1 + \exp[-(E_f - E_v)/k_B T]\}$. We make the assumption that the electrons and holes have the same effective masses, which may be relaxed if not appropriate. Under thermal equilibrium, the Fermi energy for *n*-type TMD layer is thus $E_f - E_c = k_B T \ln[\exp(n/g_{2D} k_B T) - 1]$, and for *p*-type it is $E_v - E_f = k_B T \ln[\exp(p/g_{2D} k_B T) - 1]$.

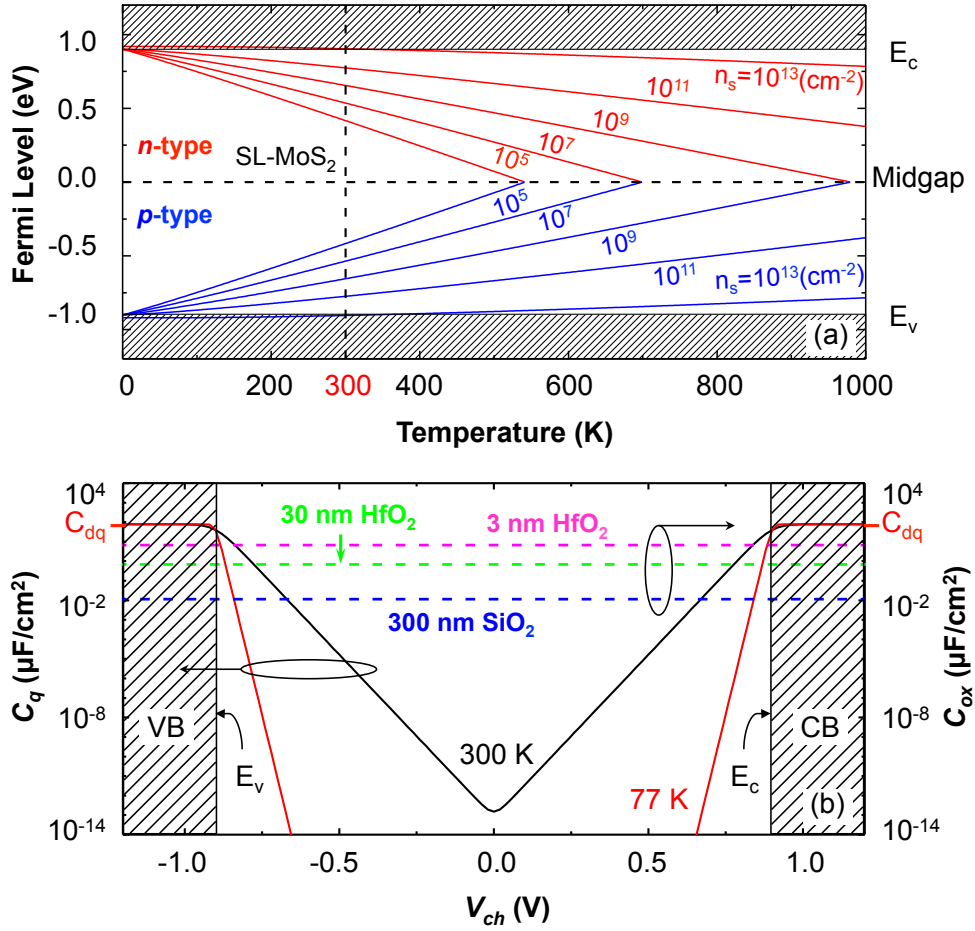

**Figure 1**. (a) Fermi level as a function of temperature for MoS$_2$ single layers for different 2D carrier densities. Red lines show Fermi levels for *n*-type and blue lines for *p*-type MoS$_2$ layers. The horizontal dashed line indicates the location of midgap and the vertical dashed line indicates the room temperature, 300 K. (b) The quantum capacitance $C_q$ as a function of the local channel electrostatic potential $V_{ch}$ at 77 K and 300 K. The electrostatic capacitances per unit area of 3 nm and 30 nm HfO$_2$, and 300 nm SiO$_2$ are shown as references. $C_{dq}$ is the degenerate limit of $C_q$.

Figure 1 (a) shows $E_f$ plotted as a function of temperature for MoS$_2$ single layers for different 2D carrier densities. The red lines are for *n*-type and the blue lines for *p*-type layers. The horizontal dashed line indicates the Fermi level for intrinsic MoS$_2$; it stays at the mid gap because of the assumed symmetric bandstructure. SL TMDs have large electron effective masses, (~0.57$m_0$ for MoS$_2$, ~0.6$m_0$ for MoSe$_2$, and ~0.61$m_0$ for MoTe$_2$).[12] As a result, the DOS is high. As shown in Fig. 1 (a), the carrier statistics stays effectively *nondegenerate* at room temperature over a very wide range of density of interest (10$^{11}$~10$^{13}$ cm$^{-2}$), with the Fermi level hardly entering the bands. As expected, at elevated temperatures the semiconductor turns intrinsic because of interband thermal excitation of carriers. The intrinsic carrier density ($n_i$) in 2D crystal semiconductors is given by

$$n_i = n = p = g_{2D} k_B T \ln\left[1 + \exp\left(-\frac{E_0}{k_B T}\right)\right], \tag{2}$$

where $E_0 = E_g/2$, $E_g$ is the band gap energy. Since in most 2D semiconductors, $E_0 \gg k_B T$,[12] $n_i$ can be approximated by $n_i \approx g_{2D} k_B T \exp(-E_g/2k_B T)$. The intrinsic sheet carrier density is low even at room temperature because of the large bandgap, for example, $n_i \sim 1.1 \times 10^{-2}$ cm$^{-2}$ for SL MoS$_2$ as compared to ~10$^{11}$ cm$^{-2}$ for zero-gap graphene.[8] The carrier density in a semiconductor cannot be lower than $n_i$ at that temperature; this is also the reason for the high achievable on-off ratios in TMD FETs compared to 2D graphene.

The effect of the gate voltage in a FET is to tune the carrier density, and consequently, the Fermi level in FET channels. A positive gate voltage applied to an intrinsic 2D crystal single layer channel populates the CB with electrons, and the Fermi level is driven from the midgap towards the CB edge. The local channel electrostatic potential $V_{ch}$, which is tuned by the gate bias, determines the electron density in the 2D crystal layer:

$$n = g_{2D} k_B T \ln\left\{1 + \exp\left[-(E_0 - qV_{ch})/k_B T\right]\right\}. \tag{3}$$

Writing the total charge density in a 2D semiconductor single layer $Q = q(p-n)$ as a function of $V_{ch}$, and using the definition of quantum capacitance $C_q = -\partial Q/\partial V_{ch}$, one obtains for 2D crystals

$$C_q = q^2 g_{2D} \left\{ \left[1+\exp\left(\frac{E_0 - qV_{ch}}{k_B T}\right)\right]^{-1} + \left[1+\exp\left(\frac{E_0 + qV_{ch}}{k_B T}\right)\right]^{-1} \right\} \approx q^2 g_{2D} \left[1 + \frac{\exp(E_g/2k_B T)}{2\cosh(qV_{ch}/k_B T)}\right]^{-1}.$$

(4)

Figure 1 (b) shows the calculated quantum capacitance for SL MoS$_2$ as a function of $V_{ch}$ at room temperature and 77 K. For intrinsic layers, $V_{ch}$ in the figure also indicates the location of the Fermi level. The electrostatic parallel-plate capacitances $C_{ox}$ (per unit area) for two dielectrics typically used as the gate oxide in TMD FETs: HfO$_2$ and SiO$_2$, are shown. Only when the Fermi level is deep inside the CB or VB, When $|qV_{ch}| > E_0$, and the quantum capacitance $C_q$ saturates and approaches the degenerate limit: $C_q \to C_{dq} = q^2 g_{2D}$. As indicated by the dielectric cases in Fig. 1 (b), for most of the nondegenerate region, $C_q$ is much lower than $C_{ox}$. For very thin dielectrics, for example: 3 nm HfO$_2$, even the degenerate limit $C_{dq}$ is comparable with $C_{ox}$. Thus the quantum capacitance can significantly influence the field effect. Device models should include $C_q$ in order to properly capture the device behavior, especially in the subthreshold region and for devices with high-κ or thin dielectrics.

When the quantum capacitance is taken into consideration, a part of the gate voltage is dropped in the channel to populate it with an electron (hole) density $n_{ch}$ ($p_{ch}$), as shown in the equivalent circuit in the inset of Fig. 2 (a). For FETs with intrinsic 2D semiconductor channels, under positive gate bias, the relationship between $V_{gs}$ and $n_{ch}$ is

$$V_{gs} = \underbrace{V_0 + V_T \ln\left[\exp\left(\frac{n_{ch}}{g_{2D}k_B T}\right) - 1\right]}_{V_{ch}} + V_{ox}, \tag{5}$$

where $V_{ch}$ and $V_{ox}$ denote the voltage drops in the channel and the dielectric layer respectively, and $V_0 = E_0/q$, $V_T = k_B T/q$ and $V_{ox} = qn_{ch}/C_{ox}$. Eq. (5) is a transcendental equation, which can only be solved numerically. The resulting $n_{ch}$ in an intrinsic SL MoS$_2$ channel as a function of $V_{gs}$ from Eq. (5) is shown in Fig. 2 (a) as black lines for 3 nm and 300 nm SiO$_2$ gate oxide. Electron densities calculated with Eq. (1) are also shown in Fig. 2 (a) as reference with blue lines. The shaded areas and the arrows indicate the error between $n_{ox}$ and $n_{ch}$. It is obvious that the carrier density can be strongly overestimated by using the commonly used expression Eq. (1) for $n_{ox}$. The large deviation proves that neglecting the quantum capacitance will lead to *significant* errors in the extraction of the carrier density.

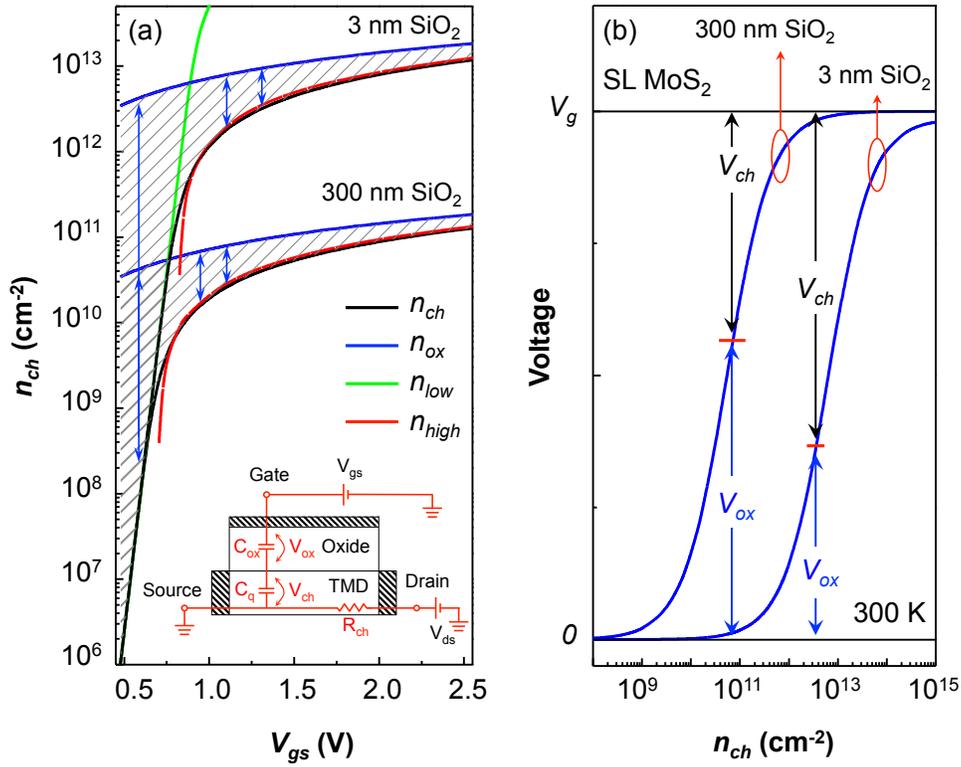

**Figure 2**. (a) Electron densities as a function of gate voltage: $n_{ch}$ is the accurate electron density calculated with the transcendental equation Eq. (5); $n_{ox}$ is the electron density obtained from Eq. (1); $n_{low}$ and $n_{high}$ are the approximated solutions to Eq. (5) at low and high gate bias respectively. The shaded areas and the arrows indicate the error between $n_{ch}$ and $n_{ox}$. The equivalent circuit of the device is shown in the inset. (b) The proportions of $V_{ch}$ and $V_{ox}$ in $V_g$ as a function of $n_{ch}$ for SL MoS$_2$ FETs coating with 3 nm and 300 nm SiO$_2$ gate dielectrics.

Reducing Eq. (5) from the transcendental form under common device operation conditions will enable the direct calculation of $n_{ch}$. At low gate voltages in the sub-threshold region of a FET where $C_q \ll C_{ox}$, most of the gate voltage drops in the channel, that is $V_{gs} \approx V_{ch}$. In this case, the electron density in the channel $n_{low}$ reduces to

$$n_{low} \approx g_{2D} k_B T \ln\left[\exp\left(\frac{V_{gs} - V_0}{V_T}\right) + 1\right], \tag{6}$$

as shown by the green line in Fig. 2 (a). $n_{low}$ arises solely due to the channel material itself, thus is independent of the gate oxide. At high gate voltages when the FET is 'strongly on', $C_q$ reaches $C_{dq}$, the channel electron density $n_{high}$ is approximately

$$n_{high} \approx \frac{1}{q} \frac{C_{ox} C_{dq}}{C_{ox} + C_{dq}} (V_{gs} - V_{cr}), \tag{7}$$

as shown by the red lines in Fig. 2 (a). $V_{cr}$ is the critical gate voltage that differentiates the situations described by Eq. (6) and (7), which corresponds to the gate voltage when $C_q = C_{ox}$,

$$V_{cr} = V_0 + V_T \ln\left(\frac{C_{ox}}{C_{dq} - C_{ox}}\right) + V_T \frac{C_{dq}}{C_{ox}} \ln\left(\frac{C_{dq}}{C_{dq} - C_{ox}}\right), \tag{8}$$

When $V_{gs} < V_{cr}$, $n_{ch}$ is determined by Eq. (6); when $V_{gs} > V_{cr}$, $n_{ch}$ is determined by Eq. (7). The critical carrier density $n_{cr}$ corresponding to $V_{cr}$ is

$$n_{cr} = \frac{C_{dq} V_T}{q} \ln\left( \frac{C_{dq}}{C_{dq} - C_{ox}} \right). \qquad (9)$$

For SL MoS$_2$ FETs with 300 nm SiO$_2$ gate oxide, $V_{cr} \sim 0.698$ V and $n_{cr} \sim 1.86 \times 10^9$ cm$^{-2}$; for 3 nm SiO$_2$, $V_{cr} \sim 0.818$ V and $n_{cr} \sim 1.87 \times 10^{11}$ cm$^{-2}$. It is worth noting that Eqs. (3) – (8) are obtained based on the intrinsic material and the assumption of zero flat-band voltage, that is, $V_{th} = V_{cr}$. If a SL MoS$_2$ is unintentionally doped with *n*-type impurities (which is typical till date), $V_{th}$ shifts by several tens of Volts toward negative values depending on the impurity density and the gate barrier thickness. In this case, the gate voltage term $V_{gs}$ in Eq. (6) and (7) should be replaced by $V_{gs} + V_{cr} - V_{th}$.

Now we discuss the validity of using Eq. (1) to estimate the carrier density in the 2D crystal FET channel. Because Eq. (1) is valid only when $V_{ox} \approx V_{gs}$, we show the proportions of $V_{ch}$ and $V_{ox}$ in $V_{gs}$ as a function of $n_{ch}$ obtained from Eq. (5) for SL MoS$_2$ FETs with 3 nm and 300 nm SiO$_2$ dielectric layers in Fig. 2 (b). As can be observed, for FET with 300 nm SiO$_2$ dielectric layer, $n_{ch}$ ranging from $10^{11} \sim 10^{13}$ cm$^{-2}$ can easily be overestimated by Eq. (1) because $V_{ox}$ is *significantly* smaller than $V_{gs}$. For the very thin 3 nm SiO$_2$ gate oxide, $n_{ch}$ can be strongly overestimated over the whole carrier density range of interest: $10^{11} \sim 10^{13}$ cm$^{-2}$, as also shown in Fig. 2 (b). For thin gate barriers, a significant amount of voltage is dropped in the semiconductor channel because of the carrier statistics, and its neglect can cause large errors.

With the correct carrier statistics, we now re-examine the methods employed to extract other important parameters from the device characteristics, for example, the carrier mobility. A commonly used method to estimate the carrier mobility in the channel is the field-effect mobility $\mu_{FE}$, given by:[9,13-17]

$$\mu_{FE} = \frac{d\sigma}{dV_{gs}}\left(\frac{1}{C_{ox}}\right) = \frac{dI_d}{dV_{gs}}\left(\frac{L}{WC_{ox}V_{ds}}\right), \quad (10)$$

where $\sigma$ is the electronic conductivity in the channel, $I_d$ is the drain current, $V_{ds}$ is the drain voltage, and $L$ and $W$ are the length and width of the channel respectively. Eq. (10) is widely used in device analysis of Si-based MOSFETs and III-V semiconductor-based FETs. However its validity in TMD devices must be re-examined. Equation (10) is derived from the fundamental drift current equation of an FET in the linear regime at small drain voltages:

$$I_d = Wqn_{ch}v_d = q\frac{W}{L}n_{ch}V_{ds}\mu_d, \quad (11)$$

where $v_d$ and $\mu_d$ are the carrier drift velocity and drift mobility in the channel respectively. To obtain Eq. (10) from Eq. (11), the first assumption is that the carrier density in the channel can be calculated using Eq. (1). For on-state device operation where $V_{gs} \gg V_{th}$, Eq. (7) captures the carrier statistics and quantum capacitance more accurately. The term $V_{cr}$ or $V_{th}$ can be eliminated by taking the derivative of $I_d$ vs. $V_{gs}$. Eq. (10) can be recast as

$$\mu_{FE} = \frac{dI_d}{dV_{gs}}\left(\frac{L}{W}\right)\frac{1}{V_{ds}}\frac{C_{ox}+C_{dq}}{C_{ox}C_{dq}}, \quad (12)$$

which amounts to replacing $C_{ox} \rightarrow C_{ox}C_{dq}/C_{ox}+C_{dq}$, which is not a fundamental new result in itself, but we emphasize that not doing so can cause significant errors. However, another implicit but more important assumption in Eqs. (10) and (12), which is barely discussed, is that the carrier mobility $\mu_d$ in the channel does not change when gate bias is varying. The derivative in Eqs. (10) and (12) can lead to significant errors when $\mu_d$ is $V_{gs}$ dependent, as we now discuss.

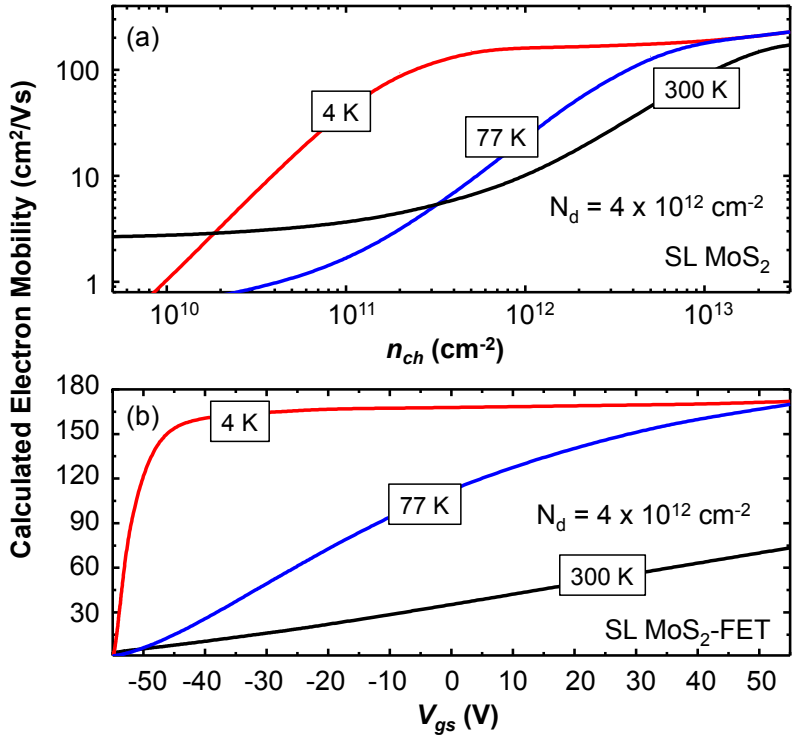

**Figure 3**. Calculated electron drift mobilities at three temperatures: 4 K, 77 K, and 300 K, as a function of (a) carrier density and (b) gate voltage.

Because the carrier density is modulated by the gate bias, the $V_{gs}$-dependence of $\mu_d$ is determined by the dependence of $\mu_d$ on the carrier density $n_{ch}$. Figure 3 (a) shows the calculated electron drift mobility in SL MoS$_2$ as a function of electron density at three different temperatures: 4 K, 77 K and 300 K. The gate dielectric is chosen as 300 nm SiO$_2$. The mobility is calculated in the relaxation-time approximation (RTA) of the Boltzmann Transport Equation (BTE). Scatterings by polar optical phonons, deformation potential phonons (acoustic and optical), remote optical phonons from the dielectric layer, and ionized impurities have been taken into consideration. Details of the calculation can be found in Ref. [3]. As can be seen from Fig. 3 (a), at all three temperatures, $\mu_d$ first increases with $n_{ch}$ and then tends to saturate at high density. At high temperature, a higher carrier density is required to fully screen Coulombic scattering potentials. For example, $\mu_d$ starts to saturate at $\sim 3\times10^{13}$ cm$^{-2}$ at 300 K, but at $\sim 4\times10^{11}$ cm$^{-2}$ for very low temperature 4 K. Combining the results of Fig. 3 (a) and Eq. (5), one can obtain the electron mobility as a function of $V_{gs}$, as shown in Fig. 3 (b). An ionized impurity density $N_d$ of $4\times10^{12}$ cm$^{-2}$ is assumed to be located in the channel, which leads to a negative shift of the threshold voltage of $\sim 55$ V from the intrinsic case based on the following relationship: $N_d \approx \left(C_{ox}^{-1}+C_{dq}^{-1}\right)^{-1}(V_{cr}-V_{th})/q$. At 4 K, the mobility starts to saturate at $\Delta V_{gs}(=V_{gs}-V_{th}) \sim 10$ V, while mobilities at 77 K and 300 K keep increasing even when $\Delta V_{gs}$ is well over 100 V. Note that the drift mobility $\mu_d$ discussed here differs from the Hall mobility $\mu_H$ by a *Hall factor*, which is induced by the magnetic field in the Hall-effect measurement. The Hall factor is often assumed to be unity, however careful consideration of the Hall factor with relevant scattering mechanisms at different temperatures needs further detailed study.[18] Baugher et al.[19] have compared $\mu_{FE}$ and $\mu_H$ and found that $\mu_{FE}$ can differ significantly from $\mu_H$. They attributed the lower $\mu_H$ to the possible screening of charged impurity scattering at higher densities, which is consistent with our results in Fig. 3. In the following, we quantitatively explain the discrepancy between the conventional method of extracting the field-effect mobility $\mu_{FE}$ and the 'true' drift mobility $\mu_d$ in the channel by combining a theoretical transport calculation with density-dependent mobility, and with the correct electrostatics of the FET incorporating the correct carrier statistics and quantum capacitance. This final analysis explains the measured experimental behavior of SL TMD FET, and highlights the problems with conventional models of mobility extraction.

Figure 4 (a) shows the experimentally obtained output characteristics (open squares) at gate voltages of 40 V, 0 V and -40 V of a typical back-gated SL MoS$_2$ FET with a 300 nm SiO$_2$ layer as the gate oxide.[15] Figure 4 (b) shows the transfer characteristics of the same device in both linear and log-linear plots at a fixed drain bias of 10 mV, The effect of the contact resistance has been de-embedded by using the experimental values.[15] Here we make the assumption that the contact resistance does not change with the gate voltage. The measured room temperature data are chosen for the study here because the contact effects play a less important role at higher temperature. The length and width of the channel are 4 μm and 9.9 μm respectively. Since the drain voltage is small, the variation of the carrier density and mobility from the source to the drain is ignored. Following the compact model proposed by Jiménez,[20] the device characteristics in Fig. 4 are first modeled by assuming a constant mobility. The calculated currents are shown as solid black lines in Figs. 4 (a) and (b). The carrier statistics are obtained from Eqs. (3) - (5). As can be observed, with constant mobility, the on-state current appears to fit well for high $V_{gs}$ ~ 20 - 40 V. However, *significant* quantitative and more importantly, *qualitative* discrepancies are observed at low $V_{gs}$. On the contrary, if we fit the current at low $V_{gs}$, we would see large errors at high $V_{gs}$. Thus we remodeled the devices characteristics by taking both the carrier statistics, and the $V_{gs}$-dependence of the electron mobility into account. This calculation is shown as red lines in Fig. 4 (a) and (b). The impurity density is used as the fitting parameter, with value of ~ $4\times10^{12}$ cm$^{-2}$. The excellent fit of the $V_{gs}$-dependent $\mu_d$ model to the experimental data over several orders of magnitude change in current indicates that if we use Eq. (10) or even Eq. (12) to extract the field-effect mobility from the FET transfer characteristics, we will be in significant error. Both the quantum capacitance and the density-dependent mobility must be included for proper extraction.

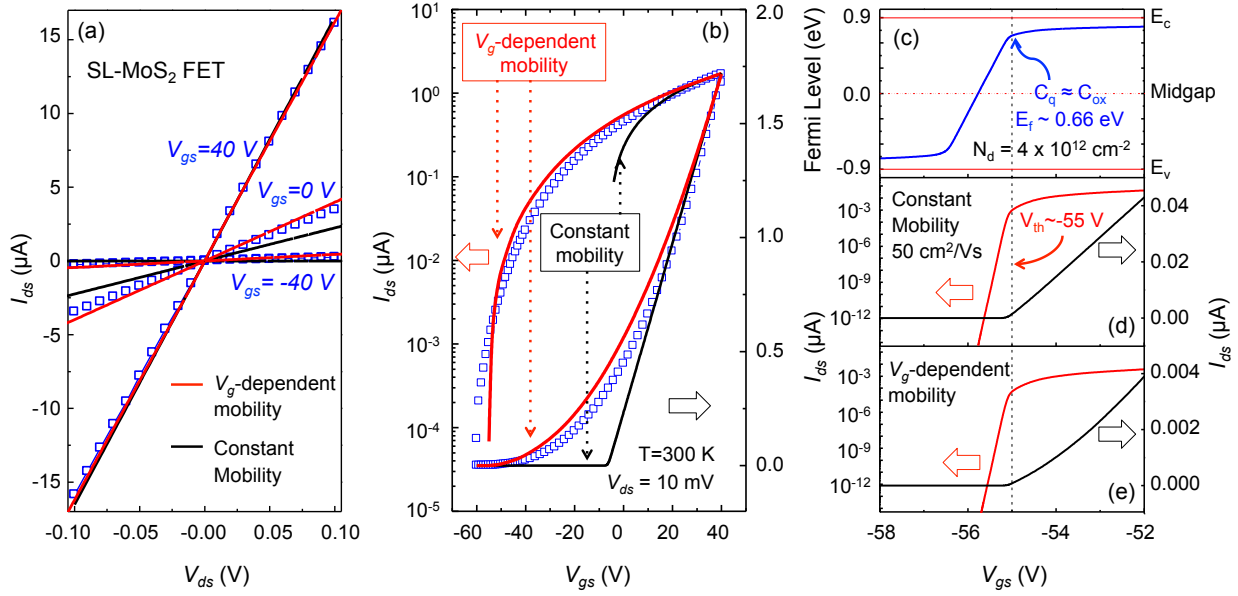

**Figure 4**. (a) Experimental output characteristics (open squares) of a typical back-gated SL MoS$_2$ FET from Ref. [15]. (b) Transfer characteristics from the same device in both linear and log-linear plots. The solid black lines show the calculated output and transfer curves with the assumption of constant electron mobility while the solid red lines are calculated with $V_{gs}$-dependent electron mobility. (c) Fermi level in the channel as a function of the gate voltage. (d) and (e) show the calculated transfer characteristics with assumed constant and $V_{gs}$-dependent electron mobility, respectively.

Figure 4 (c)-(e) show the calculated room temperature Fermi level in the SL MoS$_2$ channel, transfer characteristics with constant and $V_{gs}$-dependence mobilities respectively. The device structure is the same with that in Fig. 4 (a) and (b) and $N_d$ is fixed at $4\times10^{12}$ cm$^{-2}$. In the sub-threshold region, the drain current is dominated by the carrier density increasing with $V_{gs}$. Thus the threshold voltage $V_{th}$ can be defined as the voltage when the transfer characteristic curve has the highest curvature, as shown by the vertical dashed line in Fig. 4 (c)-(e). $V_{th}$ distinguishes the sub-threshold region and the on-state region that described by Eq. (6) and Eq. (7) respectively. For current structure, $V_{th}$ is ~ -55 V. To further prove the validity of the method of extracting $V_{th}$, we find that when $V_{gs} = V_{th}$, $E_f$ is located ~0.66 eV above the midgap, as shown in Fig. 4 (c). This is also the Fermi level when $C_q \approx C_{ox}$, as can be observed in Fig. 1 (b). Once the threshold voltage is extracted, one can now estimate the carrier drift mobility in the channel at room temperature with combining the empirical expression proposed in Ref. [3] and Eq. (7) for $n_{ch} \leq 10^{13}$ cm$^{-2}$:

$$\mu_d \approx 3500 \left( \frac{N_d}{10^{11} cm^{-2}} \right)^{-1} \left\{ A(\varepsilon_e) + \left[ \frac{1}{q} \frac{C_{ox} C_{dq} (V_{gs} - V_{th})}{(C_{ox} + C_{dq}) \cdot 10^{13} cm^{-2}} \right]^{1.2} \right\} \quad (cm^2/Vs), \qquad (13)$$

where $A(\varepsilon_e)$ is a fitting constant depending on $\varepsilon_e$, for single-gated MoS$_2$ FET with SiO$_2$ gate oxide, $A(\varepsilon_e)$ is ~ 0.036.[3]

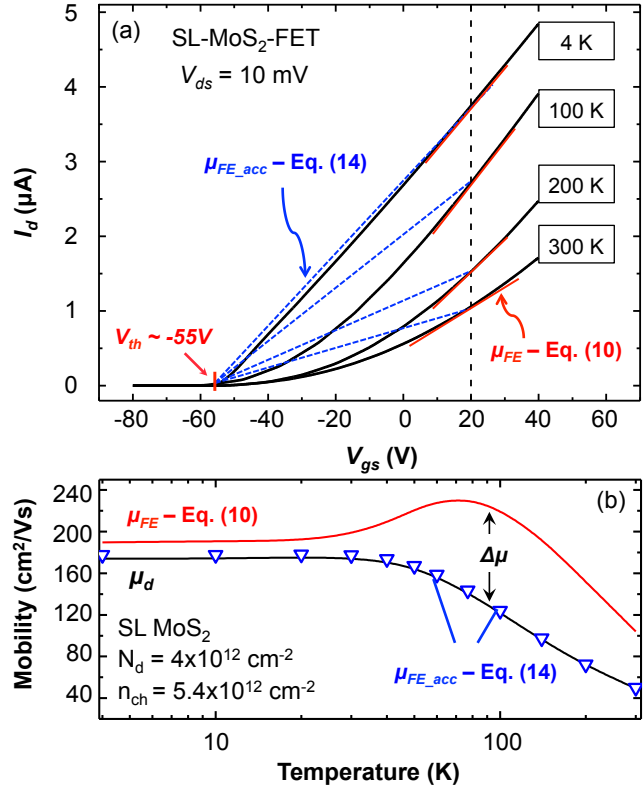

**Figure 5**. (a) Calculated transfer characteristics (black lines) of a SL-MoS$_2$ FET at temperatures 4 K, 77 K, 200 K and 300 K. The red and blue dashed lines indicate the field-effect mobility obtained from Eq. (10) and Eq. (14), respectively. (b) Field-effect mobilities at $V_{gs}$ ~ 20 V obtained from Eq. (10) and Eq. (14) as well as the drift mobility $\mu_d$ as functions of temperature.

To further show the discrepancy between the field-effect mobility and the drift mobility in the device channel, we calculate the transfer characteristics of a SL MoS$_2$ FET as a function of temperature, using the same parameters as used in Fig. 4. The example transfer curves at temperatures 4 K, 100 K, 200 K and 300 K are shown in Fig. 5 (a). Because $\mu_{FE}$ is usually extracted from the measured transfer characteristics in the region that appears to be linear,[15] for example, for $V_{gs}$ ~ 20 - 40 V in Fig. 4 (b), here we take the carrier mobility at $V_{gs}$ ~ 20 V as a case study. The carrier density at $V_{gs}$ ~ 20 V is $n_{ch}$ ~ 5.4×10$^{12}$ cm$^{-2}$. The field-effect mobilities calculated using Eq. (10) are shown by the red line in Fig. 5 (b). Because of the derivative term in Eq. (10), $\mu_{FE}$ is proportional to the slope of the tangent to the $I_d$-$V_{gs}$ curve, as indicated by the red lines in Fig. 5 (a). The black curve in Fig. 5 (b) shows $\mu_d$ calculated using our transport model. As we can see from Fig. 5 (b), $\mu_{FE}$ is higher than $\mu_d$ over the entire temperature range. Moreover, the error $\Delta\mu$ ($= \mu_{FE} - \mu_d$) is not constant as the temperature varies. The value of $\Delta\mu$ depends on the dependence of $\mu_d$ on $V_{gs}$, as was shown in Fig. 3 (b). The faster $\mu_d$ increases with $V_{gs}$, the higher is the discrepancy $\Delta\mu$. $\mu_{FE}$ calculated by Eq. (10) shows a much higher value of ~ 104 cm$^2$/Vs at 300 K while $\mu_d$ is ~ 50 cm$^2$/Vs. Conversely at 4 K, since $\mu_d$ starts to saturate at very low $\Delta V_{gs}$, $\mu_{FE}$ (~ 190 cm$^2$/Vs) is only slightly higher than $\mu_d$ (~ 175 cm$^2$/Vs). At temperature lower than 20 K, one can approximate $\mu_{FE} \approx \mu_d$ with error less than 10%. Over 20 K, $\Delta\mu$ first increases and then decreases with increasing temperature, leading to an apparent increase of $\mu_{FE}$ at temperatures ranging from ~30 K to ~80 K. This observation can partially explain the experimentally obtained decrease of the field-effect mobility as the temperature is lowered.[9] Thus we conclude that $\mu_{FE}$ extracted from the device transfer characteristics by Eq. (10) not only overestimates the electron mobility, but can also show a false temperature dependence. The red line in Fig. 5 (b) shows an anomalous increase of mobility with temperature for 30 K $<T<$ 80 K. This is not related to any real scattering mechanism, but rather has roots in using incorrect carrier statistics.

To accurately extract the carrier transport properties from the device measurements, the field-effect mobility may be obtained by:

$$\mu_{FE\_acc} = \frac{I_d}{V_{gs}-V_{th}}\left[\frac{L}{W\left(C_{ox}^{-1}+C_{dq}^{-1}\right)^{-1}V_{ds}}\right]. \qquad (14)$$

$\mu_{FE\_acc}$ extracted from the calculated transfer curves in Fig. 5 (a) using Eq. (14) are shown as open triangle symbols in Fig. 5 (b) with $V_{th}$ taken as -55 V. We can see a very good agreement between $\mu_{FE\_acc}$ and $\mu_d$. Now $\mu_{FE\_acc}$ is proportional to the slope of the straight line joining $I_d(V_{th})$ to $I_d(V_{gs}=20V)$, as indicated in Fig. 5 (a) by blue dashed lines. Comparing the slopes of the blue and red lines in Fig. 5 (a), one can easily see the error induced by Eq. (10). Note that the estimation performed here should be used under the assumption of perfect Ohmic contact (or after contact resistance has been effectively eliminated). For current TMD semiconductors, it is still a challenge to obtain Ohmic contacts with high transparency. TMD FETs with the same channel material but with different contact metals can show very different electrostatic characteristics, and thus will give false information of the channel carrier statistics and mobillities.[21-23] A number of efforts have been made to improve the contact,[16,24-28] and remarkable low contact resistances have been achieved.[29-31]

In conclusion, we have investigated the importance of the carrier statistics and quantum capacitance in understanding the characteristics of 2D crystal semiconductor electronic devices. The commonly used expressions for extracting the carrier density and field-effect mobility from the transfer characteristics of 2D semiconductor FET are demonstrated to be only valid for very limiting conditions, and prone to severe errors. By combining the correct carrier statistics, quantum capacitance, and density-dependent mobitlities, we prescribe a new method to extract the correct mobilities from the FET measurements. The results presented here are expected to be useful to place our understanding of the fundamental properties of 2D crystal semiconductors on a more firm foundation.

**Acknowledgments**

The authors thank Deep Jariwala, Dr. Vinod K. Sangwan, and Dr. Mark C. Hersam for fruitful discussions and for sharing experimental data. The research is supported in part by an NSF ECCS grant monitored by Dr. Anupama Kaul, AFOSR, and the Center for Low Energy Systems Technology (LEAST), one of the six centers supported by the STARnet phase of the Focus Center Research Program (FCRP), a Semiconductor Research Corporation program sponsored by MARCO and DARPA.